\documentclass{mn2e}

\usepackage[dvips]{graphicx}

\begin{document}

\title[Satellite Galaxies]
{Satellite number density profiles of primary galaxies in the 2dFGRS.}

\author[Sales \& Lambas]{Laura Sales, and Diego G. Lambas
\\
Grupo IATE, Observatorio Astron\'omico
de la Universidad Nacional de C\'ordoba,  Argentina.\\
Consejo Nacional de Investigaciones Cient\'{\i}ficas
y T\'ecnicas.\\
}

\maketitle

\begin{abstract}
We analyse the projected radial distribution of satellites around bright primary
galaxies in the 2dFGRS.
We have considered several primary-satellite subsamples to search for dependences
of the satellite number density profiles, $\rho(r_p)$,
on properties of satellites and primaries.
We find significant differences of the behaviour of $\rho(r_p)$ depending on 
primary characteristics.
In star-forming primaries, the satellite number density 
profile is consistent with power laws within projected distance 
$20<r_p<500$ kpc.
On the other hand, passively star-forming primaries show flat $\rho(r_p)$ for
$20<r_p\le 70$ kpc,
well fitted by generalized King models with a large core radius 
parameter ($r_c\sim 68$ kpc).
In the external regions of the haloes ($r_p>100$ kpc), the density profiles
of all primaries are well described by power laws 
$\rho(r_p)\propto r^\alpha$, although 
we notice that for red, early spectral type primaries, the
outer slope obtained is steeper ($\alpha^{red} \sim 1.12$) than that corresponding
to blue, late 
spectral type primaries ($\alpha^{blue} \sim -0.79$). 
We have tested our results by control samples of galaxies identical to the 
samples of satellites
in apparent magnitude and projected distance to the primary, but with a large relative 
velocity. 
This sample of unphysical primary-galaxy pairs shows a flat radial density profile
beyond $r_p=20$ kpc indicating that our results are not biased toward a decrease of the
true number of objects due to catalogue selection effects.
Our results can be understood in terms of dynamical friction 
and tidal stripping on satellites in the primary haloes.
These processes can effectively
transfer energy to the dark matter, flattening
the central steep profiles of the satellite distribution
in evolved systems.

\end{abstract}


\section{INTRODUCTION}
In hierarchical models of structure formation,
galaxies are mainly the result of the assembling of several small
subclumps of mass, that merge to each other into larger structures.
Nevertheless, there are also several low mass subclumps which
have survived this process, and at present time, they have
turned into satellite galaxies of the main host, the primary galaxy.
Thus, the radial distribution
of the surrounding mass, $\rho(r)$, 
can provide useful information to test current
models of formation and evolution of structure. 
 
Several analytic expressions for mass density profiles have been proposed in the literature.
Earlier on 1972, Gunn \& Gott proposed the secondary infall model to explain
the formation of objects by a continuous accretion process. 
The model
assumes an Einstein-de Sitter universe, where pressureless and
non-interacting dust particles are accreted onto an isolated massive
center. This model predicts density profiles that follow a power-law shape 
$\rho(r) \propto r^{-9/4}$.
Numerical simulations  
of the formation and non linear evolution of galaxies, 
follow mass particles from initial conditions
up to their present day configuration allowing for a detail understanding
of halo structures.
The Cold Dark Matter (CDM) hierarchical scenario have proved to be in agreement with many
cosmological observations, and has become the most widely
adopted model for structure formation.
Within this CDM scenario, a universal law was proposed by Navarro, Frenk \& White 
(1996 and 1997, NFW profile hereafter)
from their studies of dark matter haloes. 
The authors find a simple law that does not depend on either the halo mass nor 
the spectrum of initial density fluctuations and cosmological model parameters.
The NFW profile can be written as $\rho(r) \propto c/(x(1+x)^2)$, where $x$ is
the distance r normalized to the radius that contains 200 times
 the critical density of the Universe ($r_{200}$), $x=r/r_{200}$.
All the variables in this formula are related to the concentration
parameter, $c$,
which depends on the mass of the haloes.
Massive objects are characterised by lower $c$ values than the low mass systems,
so that small haloes are usually more concentrated.
Therefore, NFW dark matter haloes have a cuspy inner density profile
with $\rho(r) \propto r^{-1}$, while in the outskirts, 
the density falls as $\rho(r)\propto r^{-3}$.
This steep behaviour of $\rho(r)$ at small distances has been confirmed
by other authors (Moore et al. 1998, Klypin et al. 2001)
who find inner slopes of up to $\rho \propto r^{-1.5}$.

Observationally, at galactic scales the flat behaviour of external
HI rotational curves indicates
a linear growth of mass with galactocentric distance,
suggesting an isothermal density profile,
$\rho(r) \propto r^{-2}$. 
On the other hand, several authors 
found cores instead of cuspy behaviour at the central regions of galaxy haloes
(Salucci \& Bukert 2000, de Blok et al. 2001, Gentile et al. 2004).
CDM models are unlikely to reproduce 
such constant density profiles obtained from observational data
in the center of haloes, so alternative dark matter candidates have been proposed to
reconcile theoretical predictions with observations 
(see for instance Spergel \& Steinhardt
2000).
Nevertheless, there are some attempts to reproduce central galaxy cores in 
numerical simulations within
universe models dominated by cold dark matter.
El-Zant et al. (2001) focus their study on the 
effects of dynamical friction on density profiles of galaxies, showing 
that it is possible to obtain a final core feature in haloes starting from
a NFW cuspy initial density profile.
Within the CDM scenario, the authors explore the energy transport from
baryonic clumps to the dark matter component, which heats up the dark matter
particles and flattens the density profiles of haloes in their internal regions.

The available observational data used to determine density profiles of galactic haloes
does not extend much beyond than the optical radius, since
HI rotation curves, 
are well measured up to $r_p\sim60-100$ kpc in brightest galaxies.
At $r_p>100$ kpc, satellite galaxies represent an interesting alternative
to constrain density profiles of primary haloes starting from observations.
It is not obvious that satellites follow the mass distribution within
the parent halos and
actually, many simulation results indicates that the density profiles traced by mass particles are
steeper than that obtained using subhaloes.
Nevertheless, it is important to note that
although dark subhaloes are
poor mass tracers, galaxies may follow the mass distribution more accurately.
Astrophysical processes such as
feedback, cooling or tidal disruption may help to explain
the differences between the density profiles obtained by subhaloes and particles.
Dekel et al. 2002 have proposed that feedback processes are 
suitable mechanisms for lowering satellite densities.
A satellite with smaller density could be easily disrupted 
before being accreted onto the inner regions of the haloes,
preventing the steepening of $\rho$ at small $r$.
Also, Ghigna et al. 2001 using 
dissipationless high resolution numerical simulations
showed that, as a result of tidal disruption and mergers in cluster cores, 
subhaloes have an anti-biased
number density profile with respect to the dark matter
in the central regions of the host haloes. 
Similar results were obtained by Gao et al. (2004).
The authors implemented a semi-analytical model in high resolution simulations of cluster 
haloes finding, particularly in the inner regions, a biased distribution of dark matter
subhaloes compared to the dark matter particle distribution.
On the other hand, the authors obtain that the radial distribution of galaxies within 
simulated clusters is closely related to that of the dark matter.
In a recent work, Nagai \& Kravtsov (2004) 
also conclude that galaxies in clusters are expected to give a less biased tracer 
of the dark matter compared to dark subhaloes.
Using high-resolution numerical simulations of galaxy clusters in $\Lambda$CDM cosmology and an implementation of a hydrodynamical model for galaxy formation,
the authors find a distribution of subhaloes shallower than that of the dark 
matter particles, in agreement with previous results.
However, since tidal disruption is stronger at the 
outer regions of subhaloes, their baryonic contents
are expected to be less affected by tidal effects implying that galaxies
are better tracers of the dark matter profile. 

Satellite galaxies are used here in an attempt to put constrains on the
radial density profiles of primary galaxies.
Our work has the aim of studying the projected number density profiles of
satellites around isolated primary galaxies up to 500 kpc in the 2dFGRS.
We have searched for dependences of density profiles on several primary and satellite
properties and we have tested the results using suitable control samples.
In section 2 we describe the selection criteria adopted to construct the samples studied,
and in section 3 we show the main results obtained.
Finally, in Section 4 we discuss and summarise our principal conclusions.   

\section{PRIMARY AND SATELLITE SAMPLES}

This work is based on the 2dF final data release (Colless et al. 2001).
This survey contains almost 246000 objects, mainly galaxies; which have
been selected from the APM galaxy catalogue.
The objects are distributed covering approximately
1500 square degrees of sky in \textit{North} and \textit{South Galactic Cap}
strips as well as some random fields.
It provides  photometric and spectroscopic information, such as positions,
redshifts, accurate $b_j$, B and R SuperCosmos magnitudes,
and a spectral classification through the spectral parameter $\eta$.

We have selected two galaxy samples, one corresponding to primary galaxies
and the other to their associated satellites.
The selection criteria applied to primaries can be summarised
in a number of conditions over the redshift, luminosity and environment
of the candidate galaxies.
All our primaries have been restricted to redshifts $0.01<z<0.1$
which assures a high level of completeness in the catalogue and
the lack of corrections to the Hubble law as distance estimator.  
These primaries are brighter than $M<-18$ in the absolute blue magnitude $B_j$.
Throughout this paper we assume a Hubble constant
$H_0 = 100 Km/s Mpc^{-1}$. In order to construct a sample
with a suitable  isolation criteria for primary candidates
we require that any neighbour galaxy within a region of 700 kpc in projected
separation and $\mid \Delta V \mid\le 1000$ km/s (where $|\Delta V|$ is the line-of-sight
velocity difference) must be at least 1 magnitude fainter than the
primary candidate ($B_j^{neigh}-B_j^{prim} > 1$).
By doing so, we significantly
reduce the possibility of including a single satellite
with more than one primary. We have not imposed any further
restriction on the morphological types of the primaries since we aim
to explore possible dependences of satellite properties on
the associated primary.

Concerning the sample of satellites, we have considered
objects within projected separations of
500 kpc of any given primary, and within the range $\mid\Delta V\mid  < 500 $ Km/s.
Satellite candidates should be at least two magnitudes
 fainter than the primary, and additionally they should be
 objects fainter than $B_j = -18.5$.
The conditions applied here are similar to those in previous works
and, in our case, they provide a large number of objects suitable for
statistical studies. 

We have applied a final (conservative) restriction to this sample
 with the aim of removing
 group-like systems, then we keep only those
 primary-satellite systems with
 up to 4 satellites per central galaxy.   

With the criteria described above, the final sample of isolated primary
has 1766 objects and the satellite sample is composed by 2590 galaxies.

\section{ANALYSIS}

At large galactocentric distances, 
i.e. at more than two
optical galactic radii,
satellites can 
provide important information about the dark halo properties.
There are several works in the literature based on 
this assumption (see for instance 
Zaritsky et al. 1993 and 1997, Mc Kay et al. 2002, Prada et al. 2003).
The authors apply a statistical approach to the problem
since
primary galaxies 
outside the Local Group, have a low number of detected 
satellites per galaxy
(in our sample the average value of satellites per primary is 
$\sim$1.7).
Then, the general procedure adopted is to compose a characteristic
primary halo by considering all satellites as belonging to one
typical primary.

In our work we adopt this treatment of the data,
building several ensembles of satellites
corresponding to primaries with different characteristics
(luminosity, color index and spectral type) in order 
to explore the density profile of
luminous matter in the outer regions of haloes
 (up to 500 kpc) as well as possible
 dependencies on primary and satellite properties.

We have computed the mean projected satellite
 number density profile,
 $\rho(r_p)$, for the different samples by counting the
number of satellites in bins of projected galactocentric
distance $r_p$, 
and normalizing to the total number of primaries in each sample.
Several profiles for the baryonic component have been proposed in the literature,
with and without a central core. 
In our analysis, we have adopted two theoretical profile fits for the
observed number densities.
The generalized King profile (King, 1962) 
which takes the form 
$$\rho(r_p)=\sigma_0 \left( \frac {1}{1+{\left( \frac {r_p} {r_c}\right)} ^2}\right) ^\beta $$ 
where $\sigma_0$ is a parameter that fixes the amplitude  
and the shape of the profile is determined by 
the core radius $r_c$ and the $\beta$ parameter.
This profile was originally aimed to fit the distribution 
of stars in globular clusters
and also provides an accurate fit
to the observed radial distribution of galaxies in rich Abell clusters
(Adami et al. 1998).
Alternatively, we also consider a simple power-law fit $\rho(r_p)=Ar^\alpha$. 
 
\begin{center}
\begin{figure}
\includegraphics[width=84mm]{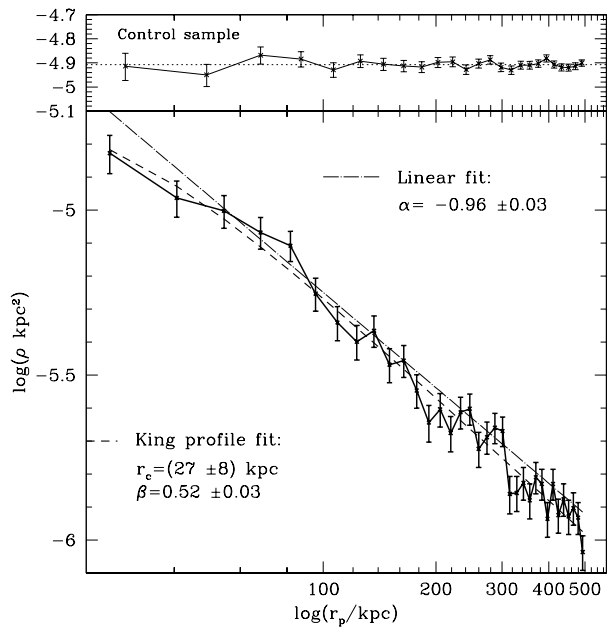}
\caption{Projected satellite number density profile in the
total sample. 
The short dashed curve indicates the generalized King profile fitted.
The best power-law fit is shown in long dashed-pointed line.
The upper panel shows in solid curve $\rho(r_p)$ corresponding to the control sample
composed by physically unbound galaxies
and in dotted line their mean value
(see text for details).
The error bars are Poisson uncertainties.}
\label{roall}
\end{figure}
\end{center}

Firstly, we have computed the satellite number density 
profile corresponding to the complete sample of primaries 
and with no restriction to satellite properties.
The results are displayed in figure \ref{roall} where we show   
the generalized King model that best fits the data with parameters 
 $r_c=28 \pm 8$ kpc
 and $\beta=0.52 \pm 0.06$. 
We also show 
a power-law fit, $\rho(r_p) \propto r_p^\alpha$ with 
$\alpha=-0.96 \pm 0.03$ consistent with
a projected isothermal halo. 
The profiles were computed by weighting each data point by its 
Poisson  uncertainty.
We have used an implementation of the nonlinear 
least-squares Marquardt-Levenberg
algorithm, and the uncertainty of the parameters
were derived from 20 bootstrap re-samplings of the data. 

In a recent work, van den Bosch et al. (2004b) analysed the abundance and radial
distribution of satellite galaxies in mock galaxy redshift surveys.
The authors find a significantly lower 
number of satellites in 2dFGRS at small separation of primary galaxies 
compared to the results of the mock catalogues.
It is argued that this deficiency could be due to 
overlap and merging of images corresponding to close projected 
galaxy pairs in the APM catalogue.
We have accounted for this possible bias in our data in the following way:
we selected a control sample of projected primary-object systems, namely,
pairs of primary-object galaxies that are close only by projection  
and not in 3-dimensional space.
This was accomplished by searching for galaxies inside 500 kpc of projected
distance from each isolated primary and at least two magnitudes fainter than it, but
with a relative difference velocity $2000<|\Delta V|<10000$ km/s.
We calculated $\rho(r_p)$ of primaries with these control objects, where
we expect the flat profile characteristic of unphysical pair systems 
in the entire range of projected separations.
We found that excluding an inner region of $r_p \sim 20$ kpc, $\rho(r_p)$
for this control sample is flat within errors (see upper panel of
fig. \ref{roall}). 
With this sample of unbound objects 
we can obtain a direct measure of how observational effects could affect the determination of
the number density
profile of isolated primary galaxies, since all physical properties (except, by definition,
the large relative velocity) of this control sample are
comparable to those of the satellites. In particular, 
we notice that the apparent magnitude distribution of control and satellite samples are 
very similar, expanding the range $16<b_j<19$. 
The flat behaviour of $\rho(r_p)$ obtained for the unbound primary-object systems provides 
a very strong argument against a significant lack of
pairs due to merge and overlap of images in APM plates and by fiber-fiber
collisions in 2dFGRS beyond $r_p=20$ kpc.
Then, according to our selection criteria, the 2dFGRS provide information
suitable to trace the number density profiles of primaries in the region
$20$ kpc$<r_p<500$ kpc without significant loss of true satellites.
The profiles and fitting parameters shown through this paper were calculated
within this interval of $r_p$.

From fig. \ref{roall} it is evident that
both analytic approximations,
King and power-law profiles, fit well
the external region of the primary haloes.
However, there is a tendency of the power-law fits to overestimate 
the amplitude of the density profile in the inner region ($r_p<50$ kpc). 

\begin{center}
\begin{figure}
\includegraphics[width=84mm]{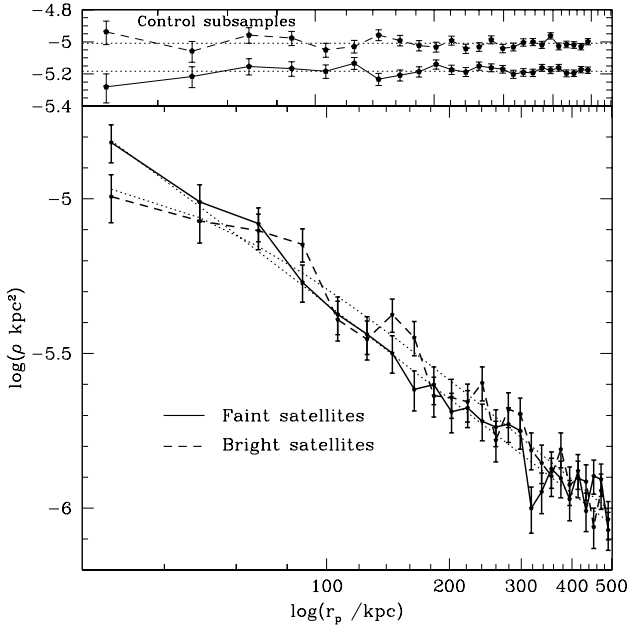}
\caption{Projected satellite number density profile of faint and bright satellite galaxies
(solid and dashed lines respectively).
In upper panel we show $\rho(r_p)$ for the equivalent control subsamples.
The error bars correspond to Poisson uncertainties.}
\label{lumsat}
\end{figure}
\end{center}

Our sample of satellites is composed of objects spanning a wide range of
absolute magnitude, color indexes and spectral types
and it is not obvious that the satellite distribution has a radial profile
independent of these properties.
For instance, an analysis of numerical simulations 
 of Taylor et al. (2003)
suggests that in galactic haloes, low-mass clumps can be found  
at smaller distances to the center than the most massive ones. 
Therefore, we may expect to find a difference in the slope of the 
profile traced by bright or faint satellites, in particular
at small separations from the primary.

In order to explore this possibility, we have divided 
the sample of satellites into bright and faint objects,
where the threshold was taken as 
the median of the absolute magnitude distribution of satellites.
The resulting profiles 
of bright and faint satellites can be appreciated
 in figure \ref{lumsat}, where solid and dashed lines 
correspond to faint ($B_j^{sat}>-17.3$)
 and bright ($B_j^{sat}<-17.3$) satellites respectively.
Based on the shape of the total sample $\rho(r_p)$, we have fitted
generalized King models to both subsamples (dotted lines). 

Consistent with the results from numerical
simulations, bright satellites 
at small projected separations from the primary, $r_p<50$ kpc, 
have a flat profile which gradually steepens 
and becomes consistent with the faint satellite profile beyond 
$r_p>100$ kpc.
It is clear from this plot that
in the external regions of haloes, bright and faint satellites
trace the same profile, but that closer to the primary,
there is a significant 
relative excess of faint satellites.
This should be reflected in the King profile fitting
parameters mainly through the core radius,
which is expected to have large values for the bright satellite radial
distribution.
The resulting parameters are 
$r_c = 48 \pm 15$ kpc and $\beta = 0.55 \pm 0.04$ for bright satellites,
and $r_c = 7 \pm 7$ kpc and $\beta = 0.51 \pm 0.03$ for the faint ones.

We have also explored the dependence of the number density profiles on 
satellite colour index and spectral type.
We have divided the sample of satellites into red and blue
objects (taking the median value of 
the satellite color index distribution as a threshold to divide the samples).
We find that
red satellites are distributed with a significantly larger core radius compared to
the blue ones.
Similarly, we divided the sample of satellites taking into account their spectral types.
Here, the threshold was taken at $\eta^{sat}=1.1$
which divides the total sample of satellites into approximately equal number subsamples.
We find that the subsample of satellites with a stronger level of
star formation activity ($\eta^{sat}>1.1$) are more frequent at
small $r_p$ values than poor star-forming satellites,
which show a flatter inner profile.

Nevertheless, 
the criteria adopted to select the 
satellite subsamples imply correlations
between primary and satellite properties.
For example 
a bright primary is likely to have
brighter satellites than a faint primary, and 
as consequence,
color indexes and spectral types  
of samples of primaries and  satellites can be strongly correlated.
Therefore it is important to consider the dependence of
the profiles on primary properties as well. 

Lorrimer et al. (1994) found that
primaries of early morphological type
tend to have their satellites more centrally clustered
than late-type primaries. 
The authors fitted power-law profiles,
$\rho(r_p) \propto r_p^\alpha$,
with values of $\alpha \sim -1.0$
and $\alpha \sim -0.8$ for early and late-type 
primaries respectively.
Therefore, we have explored 
projected satellite number density profiles around
primary galaxies with different colour indexes and spectral types 
(both properties closely related to the 
morphological type).
We have divided the total primary sample into roughly equal number 
red ($(B_j-R)>1.11$) and blue ($(B_j-R)<1.11$) primaries 
and we compute 
$\rho(r_p)$ of satellites for both subsamples.
For blue primaries,
 we find the King fitting parameters
$r_c = 1 \pm 3$ and $\beta = 0.49 \pm 0.03$.
For red primaries 
the corresponding values are $r_c = 68 \pm 12$
 and $\beta = 0.61 \pm 0.06$.
These results are in apparent disagreement with Lorrimer et al.
since red galaxies (which include more early type morphologies)
 have a larger core radius.
However, as will be discussed later, red primaries have steeper profiles in
the outer halo regions.

\begin{center}
\begin{figure}
\includegraphics[width=84mm]{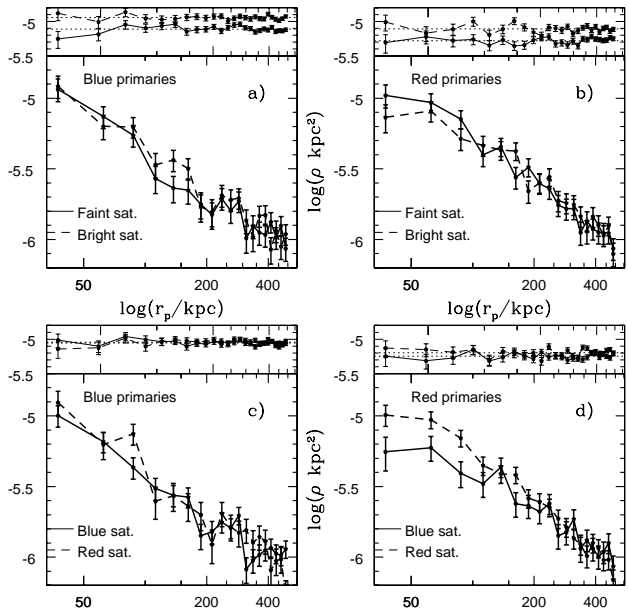}
\caption{Projected satellite number density profile for blue primaries
 (both left panels) and red primaries (right panels).
The two upper boxes show bright (dashed line) and 
faint (solid line) satellites.
In c$)$ and d$)$ we show in dashed curve the red satellites,
and in solid line the blue ones.
As before, the results of control subsamples are shown in small panels.
 The errors correspond to Poisson uncertainties}
\label{rhocolor}
\end{figure}
\end{center}

We notice that since
red primaries are brighter (with mean value of $B_j = -20.1$ compared to
the corresponding $B_j=-19.6$ of blue primaries) 
a larger fraction of bright, red associated satellites are expected
which would tend to have a large
core radius.
In order to disentangle the effects of primary and satellite 
properties on the shape of the profiles
we have considered two primary subsamples (red and blue) separately,
 and we have computed $\rho(r_p)$ for different luminosity and colour
 of satellites in each subsample of primaries. Thus, the  
result of this test could provide unambiguous evidence that
the  profiles depend mainly on 
 primary properties 
or alternatively on satellite properties. 
The results of this analysis are shown in figure \ref{rhocolor}
where the left panels ($a$) and ($c$) show profiles corresponding to
blue primaries. In the right panels ($c$) and ($d$) we show
$\rho(r_p)$ for red primaries.
We use different lines to distinguish between satellite 
luminosities (upper panels) and satellite color indexes 
(bottom panels).
By comparing left and right panels we can see that red
primary galaxies (right panels) 
have flat inner density profile,
a trend that remains unchanged
regardless of colour or luminosity of the satellites.
As a consequence, King profile fits of 
red primaries have a large core radius
($\sim 68$ kpc compared to $r_c \sim 1$ kpc for blue primaries independent
of satellite characteristics).

Due to the correlation between spectral type and colour indexes of galaxies,
we expect to find differences in density profiles of primaries
with different $\eta$ spectral types.
Madgwick et al. (2003) have proposed a classification of galaxies
based on Principal Component Analysis through a parameter $\eta$.
Taking into account this work, we have adopted 
$\eta=-1.4$ in order to divide the primary sample into 
two subsamples,
type I primaries 
(objects with $\eta^{prim}<-1.4$) and primaries with a large
star formation rate ($\eta^{prim}>-1.4$).
The satellite number density profiles obtained are consistent with
those of the previous analysis of colour indexes, in the
sense that poor star-forming primaries have larger 
values of $r_c$ and $\beta$
compared to those of active star-forming primaries.
The King parameters that best fit the data are shown in table
\ref{table1}. We also give in this table
the results of red and blue primaries and of the total
sample of primaries and satellites for a clear comparison.

\begin{table}
\caption{Parameters of the generalized King  
density profile fit for different subsamples.}
\label{table1}\center
\begin{tabular}{|c|c|c|}
\hline
\small{Subsample} & $r_c$ & $\beta$\\
\hline
\small{Total sample} & $28 \pm 8$ & $0.52 \pm 0.03$\\\\
\small{Red primaries} & $68 \pm 12$ & $0.65 \pm 0.06$\\
\small{Blue primaries}& $1 \pm 3$ & $0.48 \pm 0.03$\\\\
\small{$\eta^{prim}<-1.4$} & $56 \pm 12$ & $0.65 \pm 0.04$\\ 
\small{$\eta^{prim}>-1.4$} & $1 \pm 3$ & $0.46 \pm 0.03$ \\
\hline
\end{tabular}
\end{table} 

The flattening of $\rho(r_p)$ in the internal regions of haloes ($20<r_p<70$ kpc)
would imply a real deficiency of satellites at small $r$, 
although a possible bias due to
poor detection of satellites near bright primaries should also be considered.
Therefore in order to explore further this possibility,
we have determined $\rho(r_p)$ for faint and bright subsamples of
red and blue primaries.
We find that red primaries have larger core radii than blue primaries
for both, bright ($B_j < -20.1$) and faint ($B_j > -20.1$) galaxies.
The results for the subsample of bright primaries are $r_c^{red} = 71 \pm 15$ and 
$r_c^{blue} = 8 \pm 5$
, while for the faint primary subsample we obtained $r_c^{red} = 53 \pm 12$ 
and $r_c^{blue} = 1 \pm 5$.
Thus, we conclude that primary colour index is a main parameter that characterizes the core radius of
the radial distribution of satellites, although most luminous primaries have
larger $r_c$ once restricted to a range of colour indexes.

From inspection to figure \ref{rhocolor} we can see that
the density profile of blue galaxies is
well fitted by a power-law at all $r_p$ explored.
Red primaries however are characterized by larger values of $r_c$,
and depart significantly from a power-law at small $r_p$.
Nevertheless, in the external regions ($r_p>70$ kpc)
the satellite number density profiles are consistent with a power-law
behaviour for all samples.
A reliable estimate of the departure of $\rho(r_p)$ from a power-law
can be obtained by comparing the expected number of 
satellites, $N_e$, from a pure power-law  
to the number of objects actually found $N_{obs}$ 

We have fixed the parameters $A$ and $\alpha$ of the power-law fits $\rho(r_p)=Ar^\alpha$
for the range of distances $100<r_p<500$ kpc, where all the profiles 
are seen to be well described by power-laws.
We have considered the previously defined
red and blue, and, $\eta^{prim}<-1.4$ and $\eta^{prim}>-1.4$ primary subsamples
and we have used Monte Carlo simulations to build 
a mock halo with the same number of satellites for each of these subsamples, obeying 
a power-law consistent with the
observed parameters.
This procedure was repeated 30 times for each subsample 
to deal with statistical uncertainties.
As we must take into account the finite size of
the primaries, where no satellites are expected,
the simulated 
spherical mock haloes have been generated 
in three dimensions in the range $20<r<500$ kpc.
The generalized King model fits to these
haloes give  small core radius values, $r_c \sim 7-23$ kpc.
The comparison between $N_e$ and $N_{obs}$ 
(expected and observed number of satellites respectively) was made 
within projected distances $20<r_p<70$ kpc,
where we observe a strong flattening of density profiles and
the main results are listed in table \ref{table2}.

\begin{table}
\caption{Expected (from the best power-law fit) and observed
number of satellites, $N_e$ and $N_{obs}$ respectively,
for $20<r_p<70$ kpc.} 
\label{table2}\center
\begin{tabular}{|c|c|c|c|}
\hline
\small{Subsample} & $\alpha$ &\small{$N_e$} & \small{$N_{obs}$} \\
\hline
\small{Red primaries} &$ -1.11 \pm 0.07$ & 183 & 76 \\
\small{Blue primaries}& $-0.81 \pm 0.08$ & 93 & 109 \\\\
\small{$\eta^{prim}<-1.4$} & $-1.13 \pm 0.07$ & 202 & 96 \\ 
\small{$\eta^{prim}>-1.4$} & $-0.77 \pm 0.08$ & 87 & 93\\
\hline
\end{tabular}
\end{table}

Regarding external regions of haloes ($r_p>100$ kpc)
the values of $\alpha$ for different primary subsamples
(shown in the second column of table \ref{table2})
indicate that also at large projected distances,
$\rho(r_p)$ depends strongly on primary properties.
Red and $\eta^{prim}<-1.4$ primaries have a statistically significant
steeper outer 
density profiles than blue or $\eta^{prim}>-1.4$ galaxies.
The $\alpha$ values of poor and active star-forming primaries differ
up to a $\sim 4.5 \sigma$ level.
We have mentioned earlier that Lorrimer et al. (1994) obtained
similar results for their sample of satellites around early and late 
type primaries,
nevertheless, this
agreement with Lorrimer et al. work is present only  
when restricting our analysis to the outer $r_p>100$ kpc region.

If we consider now the inner regions of primary haloes,
table \ref{table2} suggests that
the size of the core radius of density profiles 
correlates with the deficiency of satellites near the primary
compared to the expected number from power-law fits. 
Poor star-forming primaries, which are likely to have extended core radii
show a large difference between $N_e$ and $N_{obs}$.
On the other hand, consistently with their small core radii,
this effect is not present
in active star-forming primaries.
Moreover, the situation is just the opposite, for blue or 
$\eta^{prim}>-1.4$ galaxies the number of satellites with small
$r_p$ value is larger than the corresponding to the power-law
fitted in the external regions.

\begin{center}
\begin{figure}
\includegraphics[width=84mm]{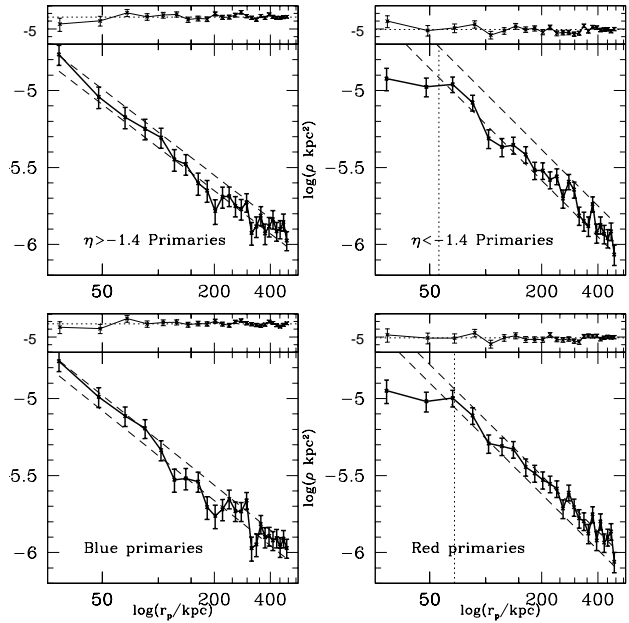}
\caption{Projected satellite number density profiles
for different primary subsamples.
Left panels show active star-forming galaxies and the right boxes 
correspond to poor star-forming primaries.
The power-law and its dispersion of simulated haloes are shown in dashed lines.
The King core radii calculated are displayed in dotted curve.
The test results obtained from control subsamples are also shown in upper small boxes.}
\label{core}
\end{figure}
\end{center}

Therefore, we can consider the core radius as suitable
parameter to indicate the scales in which the density profiles
of primary galaxies significantly depart from a power-law behaviour.   
Figure \ref{core} gives a strong support to the latter conclusion,
where we have displayed in the left panels the active star-forming galaxies
(blue primaries in the upper panel, $\eta^{prim}>-1.4$ galaxies in the bottom),
and poor star-forming primaries in the right panels
(again, upper and bottom
panels contains red and $\eta^{prim}<-1.4$ galaxies respectively).
In dashed lines we indicate the power-law density profiles of the corresponding
simulated haloes and their dispersion.
The vertical line shows the King core radius obtained for each
subsample.
In the top of all boxes we show $\rho(r_p)$ calculated using control galaxies 
associated to the same primary galaxies of each subsample.
It can be appreciated in the figures that the profiles of control samples
show no systematic decrease in the inner regions.
We conclude that the flattening of the density profile in the 
poor star-forming primary subsamples
are not likely to be caused by missing satellite images in the source catalogue, but
has instead a physical origin.

\begin{center}
\begin{figure}
\includegraphics[width=84mm]{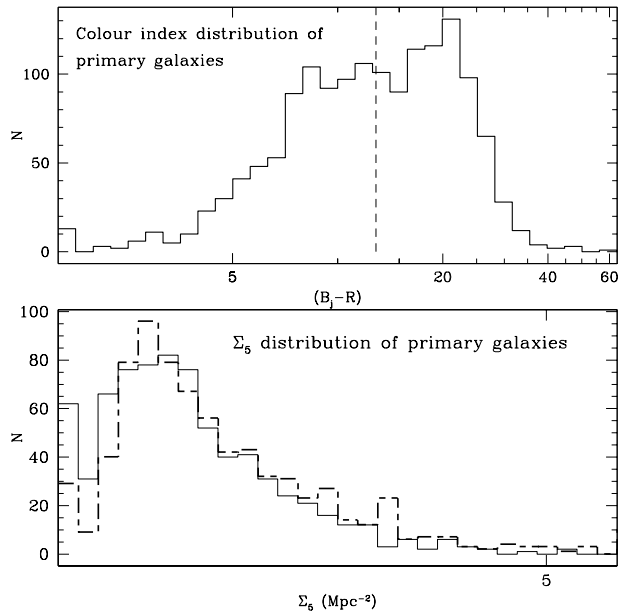}
\caption{Top box shows the $B_j-R$ colour index distribution of the primary sample.
In dashed line we show the threshold used to divide blue ($B_j-R<1.11$) 
and red ($B_j-R>1.11$) galaxies.
Bottom box shows the distribution of $\Sigma$ for these primary subsamples, solid (dashed)
line correspond to blue (red) galaxies.
} 
\label{sigma}
\end{figure}
\end{center}

Although our primary sample satisfies an isolation criteria, poor star-forming galaxies
have a higher probability of being in denser regions (probably belonging to groups or
larger systems).
This fact could increase the loss of faint satellites in primary vicinity
due to image overlap and merge.
To rule out a bias in the profiles due to differences on the primary environment,
we have explored the $\Sigma_5$ density parameter for both, red and blue
primary subsamples, obtaining a suitable statistical measure of 
how crowded is the field near each galaxy.
We calculated the projected distance from a primary to the fifth bright galaxy 
($B_j<-20$) with a relative radial velocity $|\Delta V|<1000$ km/s and estimate 
a local projected density for each primary.
The distributions of $\Sigma_5$ for red and blue primary galaxies are shown in the bottom panel
of
fig \ref{sigma}, in dashed and solid line respectively.
Both distributions are very similar, indicating that our findings are independent
of primary environment since interlopers are not likely to be different for
the two subsamples.

\begin{center}
\begin{figure}
\includegraphics[width=84mm]{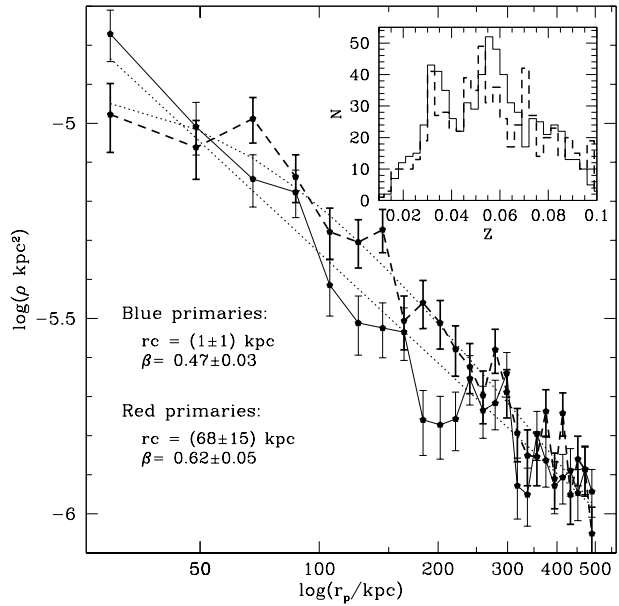}
\caption{Number of satellites density profiles for blue and red primary 
subsamples restricted to have a similar redshift distributions.
The King profiles fitted are shown in dotted curve. 
The upper small box shows the corresponding z distributions for both subsamples
(solid and dashed line respectively).
}
\label{equalz}
\end{figure}
\end{center}

We notice that due to our primary and satellite selection criteria,
bright primaries have a redshift distribution typically shifted toward
higher values than the faint subsample.
Simultaneously, because the colour-magnitude relation, 
faint galaxies tends to have smaller 
color indexes than bright ones.
Therefore, our subsample of blue primaries has smaller z values than
the red subsample, which could imply a less efficient 
detection of satellites near to red primaries.
To test this possibility, we have considered a subsample of blue and red primaries
with roughly equal z distribution (see small box of fig. \ref{equalz}).
The resulting profiles are shown in the main box of fig. \ref{equalz},
where it can be appreciated the similarity with
our previous results.
The King fitting parameters are also shown in the same figure, and are statistically
equivalent to those obtained for the full blue and red primary subsamples
(see table \ref{table1}).

Finally, we have explored for a possible dependence of the density
profiles of isolated primary galaxies on
radial velocity difference, $|\Delta V|$, and also on the number of satellite per primary.

The distribution of satellite radial velocities relative to the primary
, $|\Delta V|$,
is strongly correlated to the halo masses (Prada et al. 2003, van den Bosch et al. 2004).
The authors find that the observations are consistent with
 Gaussian distributions with dispersions increasing 
systematically with primary luminosity.
Since primary luminosity and colour are correlated,
we also expect correlations of the satellite number density profiles 
with $|\Delta V|$.
In order to explore into more detail the dependence of $\rho(r_p)$
on $|\Delta V|$,
we have repeated the previous analysis, but
considering satellites of high and low $|\Delta V|$ for our red and blue
primary subsamples.
The adopted $|\Delta V|$ threshold is 160 km/s, approximately the
mean value of the primary-satellite relative radial velocity distribution.
We find that the results previously described do not depend on $|\Delta V|$.
In both high and low $|\Delta V|$ subsamples, we find 
the satellite distribution around
red primaries to have a systematically larger core radius than around blue primaries,
as obtained before.

Our primary sample contains up to
4 satellites per system
and it is interesting to investigate if $\rho (r_p)$
depends
on the number of satellites in a system.
Thus, we considered two subsamples of primaries, the first with 1 and 2 satellites
and the second with 3 and 4.
But as before, the resulting profiles for the blue and red primary subsamples
does not depend on the number of satellites per system, with red primaries
showing flatter profiles than blue ones regardless of the number of satellite 
per primary considered.

\section{SUMMARY}

In the hierarchical scenario, the growth of structures is expected to happen
from low mass objects to larger structures. 
In such a picture, the accretion of satellites onto primaries is a common
process that may determine many properties of the large galaxies
(Klypin et al. 1999, Moore et al. 1999, Abadi et al. 2003, Moore et al. 2003).
With the aim of studying the density profile of isolated galaxies up to 500 kpc of
projected distance,
we have analysed the projected number density profiles of satellites around
bright primaries in the 2dFGRS.
Although van den Bosch et al. (2004) using mock galaxy redshift surveys
claim that 2dFGRS data is unable to
put constrains on the radial distribution of satellites around primary galaxies,
in our analysis, we have obtained flat density profiles for control subsamples 
of objects as close in projection
to primaries than real satellites and with the same apparent magnitude distribution
but with large relative velocity.
This test gives a clear indication that neither image overlap and merging of APM galaxies
nor fiber collisions
have a significant effect on our analysis of the radial distribution 
of objects near an isolated primary, 
and so our results are essentially free of such 
possible systematics in the range $20<r_p<500$ kpc.

For the total sample we find a weak, but still significant, departure 
from a power law at small galactocentric 
projected distances 
($20<r_p<50$ kpc).
However, the subsample of star-forming primaries is consistent with a pure power law fit
in the range $20<r_p<500$ kpc.
On the other hand, the satellite radial distribution of poor star-forming primaries
is characterized by a flat $\rho(r_p)$ at small $r_p$
consistent with a power law with a constant density
core. 
Accordingly, generalized King profiles with a large core radius parameter ($r_c \sim 70$ kpc)
provide good fits to the data. 
We notice, however, that $\rho(r_p)$ is significantly steeper in the outer regions
of these primaries than in blue, late spectral type primaries 
(we find $\alpha\sim -1.12$ for poor star-forming primaries and $\alpha \sim -0.79$
for galaxies with higher levels of star formation)
as it was earlier suggested by Lorrimer et al. 1994.
We have also explored dependences of $\rho(r_p)$ on satellite properties finding
that the number density profiles of faint, blue, or $\eta>1.1$ satellite subsamples 
have smaller core radii and are well fitted by power law fits.
By contrast, bright, red, or poor star-forming satellite subsamples require large
core radii.
Nevertheless, we argue that these dependencies of the number density profiles
on satellite properties are probably caused by correlations between
primary and satellite properties, since the differences between density profiles
of satellites of different types are not present when the subsamples are
restricted to belong to primaries of a given colour index and spectral type.

Results from numerical simulations can help to interpret our findings.
As pointed out in the Introduction, several authors concluded that
the cluster density profiles obtained using the radial distribution of 
subhaloes are flattened due to tidal disruption effects.
Since primary-satellites systems can be considered as
a small scale version of clusters, tidal forces
can destroy some of the satellites and bias the observed
$\rho(r_p)$.
As shown by Nagai \& Kravstov (2004) results, this tidal disruption is
expected to be stronger for subhaloes near the center of the host halo,
and as consequence, a lower number of substructures is found at small
cluster-centric distances.
Even taking into account that tidal mass loss mainly affects outer regions
of subhaloes, a percentage of baryonic mass could be loss for each satellite
orbiting in the potential well of the primary.
After several orbits , a satellite may lose a significant fraction of its
stars, falling down to the detection limit of the catalogue, or in a more 
extreme case, it can be completely disrupted by primary drag forces.
In such scenario tidal mechanisms could explain the flattening of the inner number density profile.
Nevertheless, our results indicate that only galaxies dominated by old stellar populations
(large colour index values and low $\eta$-spectral
type parameters)
have a core feature in their profiles, while for active star-forming
primaries $\rho(r_p)$ is consistent with a pure power-law.
We can interpret this behaviour as an evolutionary effect, since
primaries with a low rate of star formation 
are probably old systems so that
satellites orbiting around these galaxies 
may have been exposed to tidal forces during long time periods.

We notice that it remains unclear the validity of considering satellites galaxies as 
suitable mass tracers.
As Nagai \& Kravtsov pointed out, there are selection effects that could affect
the subhalo and the galaxy radial distribution. 
Therefore, we can not make conclusive statements on the dark matter distribution around primary galaxies
starting from their satellite radial distribution.
High resolution numerical simulations with reliable star formation schemes 
including feedback process may provide consistent models of the distributions of
mass and satellites around bright primaries and
give a detailed explanation of the observational results.

\section*{Acknowledgments}
We thank the Referee for helpful comments and suggestions which greatly improved the previous version of this paper. This work was partially supported by the
 Consejo Nacional de Investigaciones Cient\'{\i}ficas y T\'ecnicas,
Agencia de Promoci\'on de Ciencia y Tecnolog\'{\i}a,  Fundaci\'on Antorchas
 and Secretar\'{\i}a de Ciencia y
T\'ecnica de la Universidad Nacional de C\'ordoba.

\end{document}